\documentclass[final,authoryear,1p,times]{elsarticle}

 \usepackage{graphics}
\usepackage{graphicx}
 \usepackage{epsfig}

\usepackage{amssymb}

\journal{New Astronomy}

\begin{document}
\def\xxi{\xi\hspace{-0.158cm}\xi}
\def\zzeta{\zeta\hspace{-0.168cm}\zeta}
\def\al{&\!\!\!\!}
\begin{frontmatter}



\title{Oscillations of coronal loops using Rayleigh-Ritz technique}


\author[label1]{N. Fathalian}
\author[label2]{H. Safari}
\author[label1,label2]{S. Nasiri}
\address[label1]{Institute for Advanced Studies in Basic Sciences, P. O. Box 45195-1159, Zanjan, Iran}
\address{E-mail: fathalian@iasbs.ac.ir}
\address[label2]{Department of Physics, Zanjan University, P. O. Box 45195-313, Zanjan, Iran}
\begin{abstract}
In this paper, we apply the Rayleigh-Ritz Variational Scheme for
studying the transverse oscillations of a magnetic flux tube. The
flux tube is considered in low-$\beta$ solar coronal condition.
The perturbations decomposed into irrotational and solenoidal
components and MHD equations are reduced as a matrix eigenvalue
problem. In the case of longitudinally stratified thin flux tube,
 the fundamental and higher order kink frequencies are computed.
The results are in good agreements with previous studies.
\end{abstract}

\begin{keyword}
Sun: corona
\sep Sun: magnetic fields \sep Sun:
 oscillations



\end{keyword}

\end{frontmatter}


\section{Introduction}
\label{}
An understanding of the nature and propagation of waves in
magnetic structures is of considerable interest, especially as
flux tubes are important in the heating of stellar and solar
chromosphere and corona. The observational ability of spacecrafts
such as Yohkoh, SoHO, and TRACE provided us with detections of
coronal waves (e.g., Aschwanden et al. 1999; Nakariakov
et al. 1999; Wang et al. 2003; Wang \& Solanki
 2004; Berghmans \& Clette 1999; De Moortel et al. 2000).
Direct observations of fundamental quantities
such as the magnetic field strength of solar corona is still
difficult (Verth \& Erd\'{e}lyi 2008; Ruderman et al. 2008; Van
Doorsselaere et al. 2008). However, the coronal and magneto
seismology techniques allow the information to be extracted from
observations of oscillatory phenomena and the results to be interpreted
using theoretical models (Roberts et al. 1984;
 Goossens et al. 1992).

Several theoretical models of coronal loops have been developed.
The earlier models considered the simple aspects of magnetic flux
tubes (e.g. Edwin \& Roberts 1983). In their slab and cylindrical flux tube
models, infinitely long and straight uniform magnetic field,
plasma density of loop
 comparable with its environment, constant cross section,  constant
gravitation, isothermal structure, and no initial flow, were
considered.

In the recent observational data, periods, phases, damping times,
and mode profiles for coronal loops are reported by Verwichte et al.
2004; De Moortel \& Brady 2007. As
expected, the results differ from those based on simplified
theoretical models. To be more realistic several features may be
added  to this simple model, such as the presence of magnetic
twist and shells (Bennett et al. 1999; Sakai et al. 2000;
Erd\'{e}lyi \& Fedun 2006;  Erd\'{e}lyi \&
Carter 2006; Ruderman 2007), or
field-aligned flows (Terra-Homem et al. 2003), the role of line-tying effects
(D\'{i}az et al. 2004), loop curvature (Smith et al. 1997; Van
Doorsselaere et al. 2004; Brady \& Arber 2005), coronal leakage
(D\'{i}az et al. 2004; Brady \& Arber 2005).

Oscillations of longitudinally density stratified coronal loops
are investigated  by Mendoza-Brice\'{n}o et al. 2004; Andries et al.
2005a, b; Donnelly et al. 2006; Roberts 2006; Karami \& Asvar
2007; Safari et al. 2007; Andries et al. 2009; Pascoe et al.
2009. The effect of elliptical cross-section on loop oscillations
is studied by Ruderman
2003. Oscillations of  multistranded loops
 are discussed by Luna et al. 2009.
  See Aschwanden (2004, 2009) and Nakariakov \& Verwichte
(2005) for an extended review of observations of coronal
oscillations.

Despite the extensive work there are still many aspects of coronal
loop oscillations and coronal seismology to be explored. But the
previous methods, mostly based on solving differential equations,
have those own complications and inconveniences. Avoiding
complications of differential equations, a Rayleigh-Ritz
variational method can be developed. Sobouti (1981) used a
Rayleigh-Ritz variational scheme to define \emph{p}- and \emph{g}-modes
of self-gravitating fluids.
  He used a gauged version of Helmholtz's theorem
  to decompose the perturbations into an irrotational
  and a solenoidal components, which was further splitted
   into the sum of  poloidal and toroidal components.
 These components were related to \emph{p-, g-} and toroidal modes of  fluid.
Hasan \& Sobouti (1987; henceforth referred to as HS) treated
 the wave propagation in a uniform magnetic flux tube with
 a rectangular cross-section and solid boundary conditions (trapped modes).
 Nasiri (1992) extended the analysis of HS
 to include a variable cross section with vertical non-uniform magnetic filed.

Here, we investigated HS analysis to study the
transverse oscillations of a longitudinally density stratified
coronal loop. Equations of motion,
 and decomposition of Lagrangian displacements are dealt with in Sects. 2 \& 3.
 The possible motions in a cylindrical flux tube
are treated in Sec 4. The numerical results and conclusions are given in Sect.
5.
\section{Equations of motion}
The linearized MHD equations for the Eulerian perturbations of
a fluid with equilibrium values of density $\rho_0$,
gravity acceleration $g_{0}$, pressure $p_0$,
 and magnetic filed $\textbf{B}_0$ are
\begin{eqnarray}
\rho_0\frac{\partial^2\xi}{\partial t^2}=-\textbf{F}(\xi),\label{eqfz}
\end{eqnarray}
where
\begin{eqnarray}
&&\textbf{F}(\xi)=\nabla p-\rho
\textbf{g}_0-\frac{1}{4\pi}(\nabla\times\textbf{b})\times\textbf{B}_0,\\
&&\rho=-\rho_0\nabla\cdot\xi-\xi\cdot\nabla\rho_0,
\label{rho}\\&&p=-\gamma p_0\nabla\cdot\xi-\xi\cdot\nabla p_0,
\label{p}\\ &&\textbf{b}=\nabla\times(\xxi\times B_0),
\label{b}~~\textbf{v}=\frac{\partial \xxi}{\partial t},
\end{eqnarray}
in which $p$, $\rho$ and $\textbf{b}$ are the perturbations of pressure,
density and magnetic field, respectively. $\xxi({\bf x},t)$ denote a small
 Lagrangian displacement of fluid element from its equilibrium
 position. The perturbation of gravity is neglected.

 On multiplying Eq. (\ref{eqfz}) by $\xxi^*$ and integrating
over the volume initially occupied by the flux tube, one obtains
\begin{eqnarray}
\al\al\omega^2\int{d\textbf{x}\xxi^*.\rho_0\xxi}=\int{d\textbf{x}\xxi^*\cdot\textbf{F}(\xxi)}\label{eqin}\\
\al\al=\int{d\textbf{x}[\xxi^*\cdot\nabla p-\xxi^*\cdot
\textbf{g}_0 \rho-\frac{1}{4\pi}\xxi^*\cdot(\nabla\times \textbf{b})\times \textbf{B}_0]},\nonumber
\end{eqnarray}
where, $\xxi(\textbf{x},t)$ is assumed to have a time dependence proportional to $\exp(i\omega t)$.
HS have shown that, Eq. (\ref{eqin}) can be reduced as
\begin{eqnarray}
w-\omega^2s\al\al =0,\label{ws}
\end{eqnarray}
where
\begin{eqnarray}\label{s}
 s=\al\al\int{d\textbf{x}\rho_0\xxi^*\cdot\xxi},\\
\label{w}
w=\al\al \int{d\textbf{x}\frac{1}{\rho_0}\frac{dp_0}{d\rho_0}\rho^*\rho}\\
\al\al +
\int{d\textbf{x}\Gamma p_0\nabla\cdot\xxi^*\nabla\cdot\xxi}+
\frac{1}{4\pi}\int{d\textbf{x}\textbf{b}^*\cdot\textbf{b}},\nonumber
\end{eqnarray}
in which $\Gamma=\gamma-(p/\rho)(dp/d\rho)$.
From Eq. (\ref{s}), $s$ is symmetric and positive definite ($\rho_0>0$). Therefore, we may write $\omega^2=w/s$.
From Eq. (\ref{w}), the first term of $w$ is symmetric and positive ($dp_0/d\rho_0>0$).
 It can be zero if and only if $\rho=0$, which is possible if $\rho_0\xxi$ is solenoidal. The second term of
 $w$ is symmetric and positive for convectively stable fluids, i.e. when $\Gamma >0$. It is zero for convectively
 neutral fluids ($\Gamma =0$) and also for solenoidal motions($\nabla\cdot\xxi=0$).
 For convectively unstable fluids ($\Gamma<0$), the second term of $w$ contributes negatively.
 Thus, the contribution of this term to $\omega^2$ maybe positive, zero or negative.
 The last term of $w$ is symmetric and positive  too, unless $b=0$, whenever it vanishes.
 Therefore, this term also contributes positively
 to $\omega^2$. The sum $w$ is symmetric, thus ruling out either damping or overstability. There is,
 however, the possibility of dynamical instability if the fluid is convectively unstable.
   $w$ is composed of three separated terms, with distinct nature.
The first, second and third terms indicate the pressure,
buoyancy and magnetic forces, respectively.
Equation (\ref{eqfz}) or its equivalent variational form, Eq. (\ref{ws}),
constitute a generalized eigenvalue problem.
\subsection{Boundary Conditions}
\begin{figure}
       \includegraphics{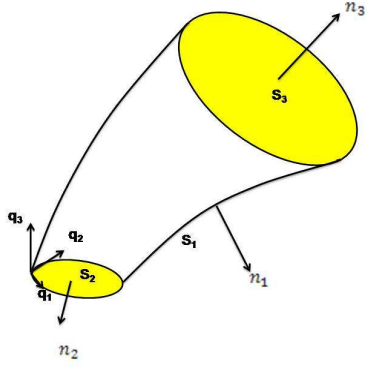}
         \vspace{4cm}
       \caption{A sketch of the equilibrium model of the flux
       tube. ${\bf n_1}$, ${\bf
n_2}$, and ${\bf n_3}$ are the normal vectors to the lateral
surface $S_1$, two end surfaces $S_2$ and $S_3$, respectively.
            }
            \label{fig1}
\end{figure}

We consider a general flux tube with its ends on the dense plasma
as shown in Fig \ref{fig1}. In the figure, ${\bf n_1}$, ${\bf
n_2}$, and ${\bf n_3}$ are the normal vectors to the lateral
surface $S_1$, and two end surfaces $S_2$ and $S_3$, respectively. The
flux coordinates, $(q_1,q_2,q_3)$, are adopted as shown in the
Fig. \ref{fig1}. We assume the hydrostatic equilibrium set of temperature,
density, gas pressure, and magnetic field,
$(T_{0i},\rho_{0i},P_{0i},{\bf B_{0i}})$ for interior, and
$(T_{0e},\rho_{0e},P_{0e},{\bf B_{0e}})$ for exterior regions of the flux tube
with an adiabatic process. The background magnetic field is
potential field and the initial material flow and the dissipative
terms are neglected. The gravity, ${\bf g_0}$, is constant.
The boundary conditions for tube surfaces which are parallel to
the equilibrium magnetic field are given as
\begin{equation}\label{b1}
\hat{n_1} \cdot [{\bf v}] = \hat{n_1} \cdot [{\bf b}]=0,~~ [P_T] =
0,\label{bound1}
\end{equation}
where $\hat{n_1}$ is the normal unit vector to the lateral
surface, $P_T=p+\frac{B^2}{8\pi}$, and $[f]=f_{\rm interior}-f_{\rm
exterior}$.  For two ends of the flux tube, the boundary
conditions are
\begin{equation}\label{b2}
 {\bf v}.\hat{n}_{2,3} = {\bf b}.\hat{n}_{2,3}= P_T= 0.\label{bound2}
\end{equation}
in which $\hat{n_2}$ and $\hat{n_3}$ are the normal unit vectors
to the end surfaces.
See also Donnelly et al. (2006) and references therein for more
details.
\section{Decomposition of Lagrangian displacements}
Let $\zzeta(\textbf{x},t)$ denote a linear displacement in the fluid.
The collection of all such displacements will belong to a Hilbert space $\cal{H}$.
We define the inner product in $\cal{H}$ of $\zzeta$ and $\zzeta'$ as
\begin{equation}\label{inner}
(\zzeta,\zzeta')=\int{d\textbf{x}\rho\zzeta^*\cdot\zzeta'};~~~
\zzeta ,\zzeta'\epsilon \cal{H}.
\end{equation}
By a suitable gauge transformation, $\zzeta$
 can be decomposed using Helmholtz's theorem as follows (see Sobouti (1981))
\begin{equation}\label{zeta}
\zzeta=\zzeta_1+ \zzeta_2+ \zzeta_3,
\end{equation}
where the various components can be expressed
in terms of scalar functions $\chi_1$, $\chi_2$, and $\chi_3$
\begin{eqnarray}
\al\al\zzeta_1=-\nabla\chi_1,\label{zeta1}~~(\rm{irrotational}),\\
\al\al\zzeta_2=\nabla\times\nabla\times(\hat{q_3}\chi_2),\label{zeta2}~~(\rm{toroidal}),\\
\al\al\zzeta_3=\nabla\times\nabla\times\nabla\times(\hat{q_3}\chi_3),\label{zeta3}~~(\rm{poloidal}).~
\end{eqnarray}
It can be shown that,
the decomposition is complete, however, not unique.
Let $\{\xxi_l;l=1,2,3\}$ be an eigenvector of Eq. (\ref{eqfz}).
Consider $\{\zzeta_s;s=1,2,3\}=\{\zzeta_{1i},\zzeta_{2j},\zzeta_{3k};i,j,k=1,2,...\}$
as a complete set of basis vectors in the Hilbert space of the displacement vectors
that satisfies equations of motion. Expanding $\xxi_l$ in terms of $\{\zzeta_s\}$, one gets
\begin{equation}
\xxi_l=\sum_s\zzeta_s Z_{sl},\label{xil}
\end{equation}
where $Z_{sl}$ are proportionality constants. Substituting Eq. (\ref{xil}) into
Eqs (\ref{ws})-(\ref{w}) and using a variational technique to
minimize the eigenvalue, gives the following matrix equation
\begin{equation}
\textbf{W}\textbf{Z}=\textbf{S}\textbf{Z}\textbf{E},\label{WS}
\end{equation}
where $\textbf{\rm{E}}$ is a diagonal matrix whose elements are the eigenvalues $\omega_l$
and $\textbf{Z}=[Z_{ls}];$ $l,s=1,2,3$, is the matrix of the expansion coefficients.
The functional expressions for the
elements of $\textbf{W}$ and  $\textbf{S}$ are the same as of Eqs (\ref{s}) and (\ref{w})
with $\xxi$ replaced by $\zzeta_s$. The functional forms used for $\zzeta_s$,
for different three displacements are given in \ref{appa1}.
\subsection{Block structure of the S, W, E and Z matrices}
Schematically, the blocks of $\textbf{S}$ have following structure
\begin{equation}
\textbf{S}=\pmatrix{
      S_{11} & S_{12}      & S_{13} \cr
      S_{21}    & S_{22} & S_{23}  \cr
      S_{31}      & S_{32}      & S_{33}}.
\end{equation}
Explicit expressions for the block elements of
$\textbf{S}$ are given in \ref{appa2}.
The block structure of $\textbf{W}$ matrix will be
\begin{equation}
\textbf{W}=\pmatrix{
      W_{11} & W_{12}      & W_{13} \cr
      W_{21}    & W_{22}      & W_{23} \cr
      W_{31}      & W_{32}      & W_{33}}.
\end{equation}
The blocks of $W_{23}$ and $W_{13}$ will be zero.
Explicit expressions for the block elements
 of $\textbf{W}$ are given in \ref{appa3}.
By definition, the matrix of eigenvalues is block diagonal,
\begin{equation}
\textbf{E}=\pmatrix{
      E_{11} & 0      & 0  \cr
      0      & E_{22} & 0  \cr
      0      & 0      & E_{33}},
\end{equation}
where each block is itself a diagonal matrix. The matrix $\textbf{Z}$
has a full structure, i.e., in general, all modes have components in all there
subspaces. Thus, the $\textbf{Z}$ matrix can schematically be written as
\begin{equation}
\textbf{Z}=\pmatrix{
      Z_{11} & Z_{12}      & Z_{13} \cr
      Z_{21}    & Z_{22}      & Z_{23} \cr
      Z_{31}      & Z_{32}      & Z_{33}}.
\end{equation}
\subsection{Method of solution}
In order to determine the frequencies, we need to solve Eq. (\ref{WS}),
which was obtained from the original Eq. (\ref{ws}) by applying a
variational principle. We adapt a Rayleigh-Ritz procedure
and approximate the linear series in Eq. (\ref{WS}) by a finite number of terms,
say $n$. The matrix blocks $S_{ls}$ and $W_{ls};l,s=1,2,3$ become $n\times n$ matrices.
\section{The case of cylindrical flux tube}
Let us consider a special circular cylinder model of coronal loop.
The cylinder is considered
 with its ends at the photosphere and with a relatively
small curvature (i.e., the radius of curvature of the loop is much
larger than the loop length) to be pervaded by a uniform magnetic
field along its axis, $B = B_0\hat{z}$, and to have low gas
pressure (low-$\beta$ approximation) with isothermal equilibrium state. The length and radius of the
loop are L and R, respectively. We adopt a cylindrical
coordinates, (r, $\phi$, z) with the $z$ axis parallel to the
magnetic field (the $\hat{q_3}$ axis in flux coordinate) and
 the origin at the center of one tube ends. The
density is assumed to be
\begin{equation}
\rho(\epsilon,z)=\left\{\begin{array}{cc}
\rho_{i}(\epsilon)f(\epsilon,z),&r\leq R,\\
\rho_{e}(\epsilon)f(\epsilon,z),&r\geq R,\end{array}\right.
\end{equation}
\begin{equation}
f(\epsilon,z)=\exp{(-\frac{\epsilon}{\pi}\sin{\frac{\pi z}{L}}});~~
\rho_{i,e}(\epsilon)=\frac{\rho_{i0,e0}}{\int^L_0{f(\epsilon,z)dz}},
\end{equation}
where $\epsilon=L/H$ ($H$ is the density scale height) and $\rho_{i0}$
and $\rho_{e0}$ are the interior and exterior densities at the
footpoints of the loop, respectively.
The proposed trial functions for $\chi_l$, are considered as
\begin{equation}\label{ansatz}
\chi_{l(l=1,2,3)}=\left\{\begin{array}{cc}
F_{i}(r)e^{im\phi}Z_l(z),&r\leq R,\\
{a_{mn}}F_{e}(r)e^{im\phi}Z_l(z),&r\geq R.\end{array}\right.
\end{equation}
In the case of the body waves, $F_{i}(r)=J_m(\gamma_{mn}^i r)$, ${\gamma_{mn}^i}^2>0$,
and the surface waves $F_{i}(r)=I_m(|\gamma_{mn}^i|r)$, ${\gamma_{mn}^i}^2<0$.
For both body and surface waves,
we consider the evanescent waves for exterior, $F_{e}(r)=K_m(\gamma_{mn}^e r)$, ${\gamma_{mn}^e}^2<0$.
In above formula, $\gamma_{mn}^{\rm i}$ and  $\gamma_{mn}^{\rm e}$ are the radial wave numbers
for interior and exterior, respectively.

To satisfy Eqs. (\ref{b1}) and (\ref{b2}),
 $Z_l(z)$ functions are chosen as
\begin{equation}\label{ansatzofz}
\begin{array}{cc}
Z_1(z)=\cos{kz},\\
Z_2(z)=\sin{kz},\\
Z_3(z)=\sin{kz}.\end{array}\\~~
k=\frac{n\pi}{L},\\~~0<z<L
\end{equation}
Using Eqs. (\ref{zeta})-(\ref{zeta3}), (\ref{ansatz}), and (\ref{ansatzofz}),  and
imposing boundary conditions (\ref{b1}) and (\ref{b2}) we obtain
\begin{equation}\label{thincond}
\frac{1}{\gamma_{mn}^{i}}\frac{J'_m(\gamma_{mn}^{i}R)}{J_m(\gamma_{mn}^{i}R)}=
\frac{1}{\gamma_{mn}^{e}}\frac{K'_m(\gamma_{mn}^{e}R)}{K_m(\gamma_{mn}^{e}R)},
\end{equation}
\begin{eqnarray}\label{amn}
a_{mn}=-\frac{{B_0^2}_i}{{B_0^2}_e}\frac{J_m(\gamma_{mn}^{i}R)}{K_m(\gamma_{mn}^{e}R)}
=-\frac{{B_0^2}_i}{{B_0^2}_e}\frac{J'_m(\gamma_{mn}^{i}R)}{K'_m(\gamma_{mn}^{e}R)},
\end{eqnarray}
where $J'_m(x)$ is $dJ_m(x)/dx$.
In thin tube approximation, $|\gamma_{mn}^{i,e} | R\ll 1$, Eq. (\ref{thincond})
leads to ${\gamma_{mn}^{i}}^2={\gamma_{mn}^{e}}^2$.
\section{Results and Conclusions}
\subsection{Results}
As typical parameters for a coronal loop, we assume $\eta$ $(=R/L)=0.01$ for the ratio of radius to length,
$\rho_{e}/\rho_{i} = 0.1$ for the ratio of interior to exterior domain of density,
$B_{i}/B_{e}=0.98$ for the ratio of interior to exterior domain of magnetic field (Aschwanden 2004).
The $\beta$ parameters are $0.035$ and $0.00035$ for the loop interior and exterior, respectively.

\begin{figure}
 \includegraphics{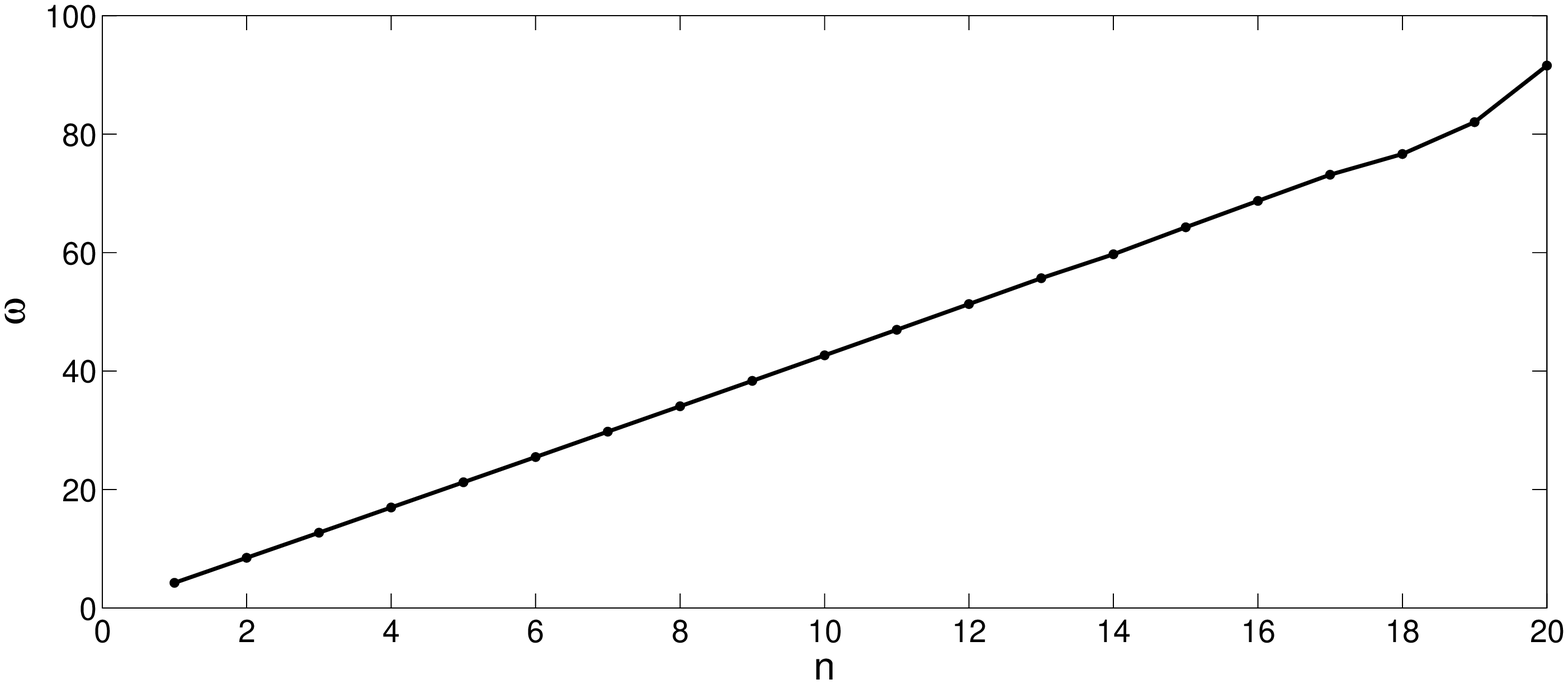}
         \vspace{7cm}
       \caption{The frequencies for unstratified loop ($\epsilon = 0$) versus longitudinal mode number, $n$.
 All frequencies are in units of $\pi v_{{A}_i}(\epsilon=0)/L$.
            }
            \label{fig2}
\end{figure}

\begin{figure}
 \includegraphics{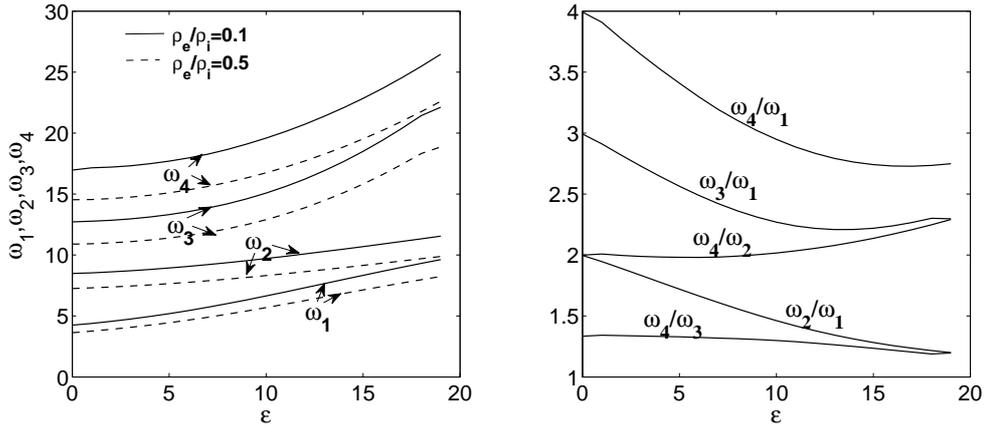}
         \vspace{4.5cm}
       \caption{Fundamental, first, second, and third overtone frequencies versus $\epsilon$.
 All frequencies are in units of $\pi v_{{A}_i}(\epsilon=0)/L$.
            }
            \label{fig34}
\end{figure}

Applying a numerical code, Eq. (\ref{WS}) is solved for eigenvalues and eigenfunctions.
For unstratified loop ($\epsilon=0$), the frequencies have been calculated and plotted versus
longitudinal mode numbers $n$ in Fig. \ref{fig2}.
As expected, $\omega_n$ is proportional to its mode number (i.e., $\omega_n\approx n\omega_1$).
For stratified loops, in a range of $0 \leq \epsilon < 20$, we have calculated the fundamental,
 first, second, and the third overtone kink ($m=1$) frequencies
$\omega_1$, $\omega_2$, $\omega_3$, and $\omega_4$, respectively. The frequencies and their ratios
are plotted in  Fig. \ref{fig34}.
As anticipated from the behavior of $f(\epsilon, z)$, all
 of the frequencies show monotonic
  increase with increasing $\epsilon$. As shown in the figure, for
small $\epsilon$, the odd modes ($\omega_1$ and $\omega_3$) has steeper slopes than
the even modes ($\omega_2$ and $\omega_4$), but modes successively  approach each
other as $\epsilon$ increases. The ratios $\omega_n/\omega_1$ ($n=2,3,4$) begins from $n$ for unstratified
loops, $\epsilon = 0$, and decreases as $\epsilon$ increases.
This result is in agreement with Safari et al. (2007). From the TRACE
data, Verwichte et al. (2004) find the ratio of the fundamental period
to the first overtone to be $1.64$ and $1.81$ for two
of their observed loops. Van Doorsselaere et al. (2007) revisited the same ratios from the observational
data to be $1.58$ and $1.82$ for the same loops, respectively, and $1.795$ for another
loop in their current analysis.
From Fig. 3.,
corresponding values of $\epsilon$ are $7.44$ and $3.69$.
Therefore, if we suppose typical loop lengths, $L = 100$ and $400 Mm$, then the
density scale heights($H$) fall in the range of $H =\epsilon^{-1}L\simeq$ [13, 53]
and [27, 108] Mm, respectively.
As you see, the ratio of the fundamental
 period to the second starts from 3 and to the third start from 4. We should note
 that, due to variational methods, the lower frequencies are more
 precise and reliable. But their total
 treatment is in agreement with previous works.

\begin{figure}
 \includegraphics{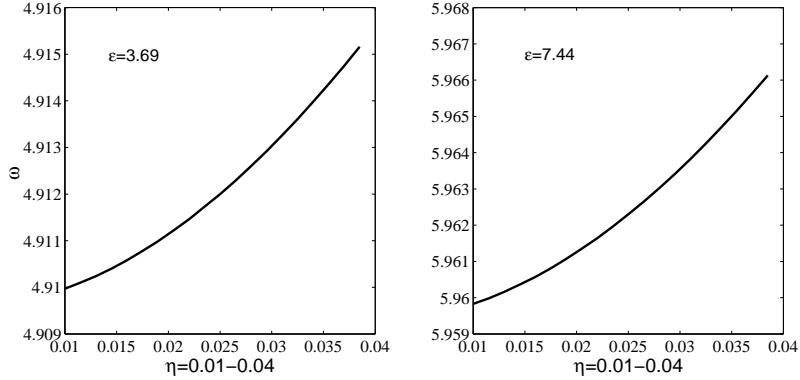}
         \vspace{4cm}
       \caption{frequencies versus $\eta=R/L$, for $\epsilon=3.69$ and $7.44$.
 All frequencies are in units of $\pi v_{{A}_i}(\epsilon=0)/L$.
            }
            \label{fig50}
\end{figure}

From the observation, we know that $\eta$
differs from 0.01 to 0.04 (e.g. Goossens et al. 2002).
 Applying this variation on the model, the frequencies show a little increase.
 The result are shown in  Fig. \ref{fig50} for two corresponding
 values of $\epsilon$.
\subsection{Conclusions}
Still many open problems remain to be solved in the field of oscillations
and  real 3D structures of the coronal loops. Actually, the real
structure of the magnetic filed, variable cross section or high
$\beta$ plasmas, yet have not been modeled completely. The
variational principles have already been applied to the problem of loop
oscillations with regard to the simplified one-dimensional
longitudinal kink equation obtained in the approximation of a "thin
tube". However, Andries et al. (2009) pointed out recently that the
variational principle may be used in a much more general sense
involving integration in both spatial directions. This
is exactly what Eq. (\ref{WS}) is about.

Here, the application of Rayleigh-Ritz variational method
to study the transverse oscillation of a general configuration of magnetic
flux tube has been investigated.
As a special case, we study the oscillations of
coronal loops with exponential density variations along the
loop axis. The kink frequencies for unstratified loops,
 and the kink frequencies and the frequency ratios for stratified loops are found.
 Increasing the density contrast decreases the frequencies but their ratios and
shape of the profiles remain unchanged. For typical loop lengths, $100-400$Mm, the density
scale heights fall in the range of $13-108$Mm, in agreement with Andries et
al. (2005a, b), Safari et al. (2007), McEwan et al. (2006), and Donnelly et
al. (2006). The frequencies show a little increase as $\eta$ increases.

So despite the fact of complication of solving differential equations,
the Rayleigh-Ritz variational method could go ahead
in a definite and obvious process and reproduce the previous results.

Choosing the trial functions for each desired
problem may be known as a challenge of this method.
Having a suitable set of trial functions,
this method could be extended to other studies of oscillations.
In the case of this work, the trial functions have a simple forms
but their selection is not so easy for more complicated geometries.
The lack of suitable trial functions affects the
convergency of the method, as well. In the case of zero external
densities, like the investigation of internal oscillations of stars,
 the trial functions could be selected such that
 the calculations become more easy and the $S$ matrix becomes diagonal.

Another problem caused by variational method is missing the precision
as one goes from the lower order modes to the higher ones. However,
the results are satisfactory enough precise
concerning the observational data in touch.

\emph{Acknowledgments:}
\small{We would like to thank Prof. Y. Sobouti for his helpful comments
through the work.}
\appendix
\section{  Displacement vectors}
\label{appa1}
Using  Eqs. (\ref{zeta1})-(\ref{zeta3}), (\ref{ansatz}) and (\ref{ansatzofz}) one gets
\begin{eqnarray}
&&\zzeta_1^{mn}=e^{im\phi}\left\{\begin{array}{cc} -\gamma_{mn}
J'_m(\gamma_{mn}r)\cos(kz)\hat{r}-\frac{im}{r}J_m(\gamma_{mn}r)\cos(kz)\hat{\phi}
+kJ_m(\gamma_{mn}r)\sin(kz)\hat{z},&r\leq R,\\{a_{mn}} (-\gamma_{mn}
K'_m(\gamma_{mn}r)\cos(kz)\hat{r}-\frac{im}{r}K_m(\gamma_{mn}r)\cos(kz)\hat{\phi}
+kK_m(\gamma_{mn}r)\sin(kz)\hat{z}),&r\geq R,\\
\end{array}\right.\nonumber\\
&&
\zzeta_2^{mn}=e^{im\phi}\left\{\begin{array}{cc} k\gamma_{mn}
J'_m(\gamma_{mn}r)\cos(kz)\hat{r}+\frac{ikm}{r}J_m(\gamma_{mn}r)\cos(kz)\hat{\phi}
+\gamma^2J_m(\gamma_{mn}r)\sin(kz)\hat{z},&r\leq R,
\\{a_{mn}}
(k\gamma_{mn}
K'_m(\gamma_{mn}r)\cos(kz)\hat{r}+\frac{imk}{r}K_m(\gamma_{mn}r)\cos(kz)\hat{\phi}
+\gamma^2K_m(\gamma_{mn}r)\sin(kz)\hat{z}),&r\geq R,\\
\end{array}\right.\nonumber\\
&&
\zzeta_3^{mn}=e^{im\phi}\left\{\begin{array}{cc}
(\gamma_{mn}^2+(n\pi)^2)[-\frac{im}{r}J_m(\gamma_{mn}r)\hat{r}+\gamma
J'_m(\gamma_{mn}r)\hat{\phi}]\sin(kz), &r\leq R,\\{a_{mn}}
((\gamma_{mn}^2+(n\pi)^2)[-\frac{im}{r}K_m(\gamma_{mn}r)\hat{r}+\gamma
K'_m(\gamma_{mn}r)\hat{\phi}]\sin(kz)),
&r\geq R.\\
\end{array}\right.\nonumber
\end{eqnarray}
\section{ Elements of S matrix}
\label{appa2}
The elements of $S$ matrix are ($m=1$)
\begin{eqnarray}
S_{11}^{nn'}&=& 2\pi(\rho_{i}(\epsilon)\rm{fR1}
+{a^*_{mn}}{a_{mn'}}\rho_{i}(\epsilon)\rm{fR2})
\times\rm{fZ1}\nonumber\\
&&+2\pi(n\pi)(n'\pi)(\rho_{i}(\epsilon) \rm{fR3}
+{a^*_{mn}}{a_{mn'}}\rho_{e}(\epsilon)\rm{fR4})
\times\rm{fZ2},\\
S_{22}^{nn'}&=& 2\pi(n\pi)(n'\pi)(\rho_{i}(\epsilon)\rm{fR1}
+{a^*_{mn}}{a_{mn'}}\rho_{i}(\epsilon)\rm{fR2})
\times\rm{fZ1}\nonumber\\
&&+2\pi\gamma_{mn}^2\gamma_{mn'}^2(\rho_{i}(\epsilon) \rm{fR3}
+{a^*_{mn}}{a_{mn'}}\rho_{e}(\epsilon)\rm{fR4})
\times\rm{fZ2},\\
S_{33}^{nn'}&=&
2\pi(\gamma_{mn}^2+(n\pi)^2)(\gamma_{mn'}^2+(n'\pi)^2)
(\rho_{i}(\epsilon)\rm{fR1}\nonumber\\
&&+{a^*_{mn}}{a_{mn'}}\rho_{e}(\epsilon)
\rm{fR2})\times\rm{fZ2},\\
S_{12}^{nn'}&=&
-\int{d S [\chi_1^*\rho\zzeta_2\cdot\hat{n}]}+\int{d\textbf{x}\frac{d\rho}{dz}
\chi_1^*{\zzeta_2}_z}
 \nonumber\\
 &=&-2\pi  n'\pi (\rho_{i}(\epsilon)\rm{fR5}+{a^*_{mn}}{a_{mn'}}\rho_{e}(\epsilon)\rm{fR6})\times\rm{fZ1}\nonumber\\
&&+2\pi\gamma_{mn'}^2(\rho_{i}(\epsilon)\rm{fR3}
+a^*_{mn}{a_{mn'}}\rho_{e}(\epsilon)\rm{fR4})\times\rm{fZ3},\\
S_{13}^{nn'}&=&
-\int{d S[\chi_1^*\rho\zzeta_3\cdot\hat{n}]}\nonumber\\
&=&
2\pi I
(\gamma_{mn'}^2+(n\pi)^2)(\rho_{i}(\epsilon)\rm{fR7}
+{a^*_{mn}}{a_{mn'}}\rho_{e}(\epsilon)\rm{fR8}) \times \rm{fZ4},\\
S_{23}^{nn'}
&=&\int{d S[\rho\frac{d\chi_2^*}{dz}\zzeta_3\cdot\hat{n}]}\nonumber\\
&=&
-2\pi I n\pi(\gamma_{mn'}^2+(n\pi)^2)(\rho_{i}(\epsilon)
\rm{fR7}+{a^*_{mn}}{a_{mn'}}\rho_{e}(\epsilon)\rm{fR8})
 \times \rm{fZ4},
\end{eqnarray}
in which
\begin{eqnarray}
\rm{fR1}&=&\int_0^1{r
dr(\frac{1}{2}\gamma_{mn}\gamma_{mn'}
[J_{m-1}(\gamma_{mn}r)J_{m-1}(\gamma_{mn'}r)+J_{m+1}(\gamma_{mn}r)J_{m+1}(\gamma_{mn'}r)])},\nonumber\\
\rm{fR2}&=&\int_1^\infty{r
dr(\frac{1}{2}\gamma_{mn}\gamma_{mn'}
[K_{m-1}(\gamma_{mn}r)K_{m-1}(\gamma_{mn'}r)+K_{m+1}(\gamma_{mn}r)K_{m+1}(\gamma_{mn'}r)])},\nonumber\\
\rm{fR3}&=&\int_0^1{r drJ_m(\gamma_{mn}r)J_m(\gamma_{mn'}r)},\nonumber\\
\rm{fR4}&=&\int_1^\infty{r drK_m(\gamma_{mn}r)K_m(\gamma_{mn'}r)}\},\nonumber\\
\rm{fR5}&=&\frac{\gamma_{mn'}}{2}J_{m}(\gamma_{mn}R)(J_{m-1}(\gamma_{mn'}R)-J_{m+1}(\gamma_{mn'}R)),\nonumber\\
\rm{fR6}&=&\frac{\gamma_{mn'}}{2}K_{m}(\gamma_{mn}R)(K_{m-1}(\gamma_{mn'}R)+K_{m+1}(\gamma_{mn'}R))],\nonumber\\
\rm{fR7}&=&\frac{\gamma_{mn'}}{2}J_{m}(\gamma_{mn}R)(J_{m+1}(\gamma_{mn'}R)+J_{m-1}(\gamma_{mn'}R)),\nonumber\\
\rm{fR8}&=&\frac{\gamma_{mn'}}{2}K_{m}(\gamma_{mn}R)(K_{m+1}(\gamma_{mn'}R)-K_{m-1}(\gamma_{mn'}R)),\nonumber\\
\rm{fZ1}&=&\int^1_0{dz f(\epsilon,z)\cos{n\pi z}\cos{n'\pi z}},\nonumber\\
\rm{fZ2}&=&\int^1_0{dz f(\epsilon,z)\sin{n\pi z}\sin{n'\pi z}},\nonumber\\
\rm{fZ3}&=&\int^1_0{dz \frac{d f(\epsilon,z)}{dz}\cos{n\pi z}\sin{n'\pi z}},\nonumber\\
\rm{fZ4}&=&\int^1_0{dz f(\epsilon,z)\cos{n\pi z}\sin{n'\pi z}},\nonumber\\
\rm{fZ5}&=&\int^1_0{dz\cos{n\pi z}\cos{n'\pi z}},\nonumber\\
\rm{fZ6}&=&\int^1_0{dz\sin{n\pi z}\sin{n'\pi z}}.\nonumber
\end{eqnarray}
\section{ Elements of W matrix}
\label{appa3}
Simplifying Eq. (\ref{w}), the elements of $W$ matrix are
\begin{eqnarray}
\frac{4\pi}{B_0^2}W_{ij}\al=\al(1+\frac{\Gamma\beta}{2})\int{
d\bf{x}\nabla\cdot\zzeta_i^*\nabla\cdot\zzeta_j}
-\int{ d\bf{x}\frac{\partial\zzeta_{zi}^*}{\partial z}\nabla\cdot\zzeta_j}\nonumber\\
\al\al-\int{
d\bf{x}\nabla\cdot\zzeta_i^*\frac{\partial\zzeta_{zj}}{\partial
z}} +\int{ d\bf{x}\frac{\partial\zzeta_i^*}{\partial
z}\cdot\frac{\partial\zzeta_j}{\partial z}},\\
\frac{2}{B_0^2}_i W_{11}^{nn'}&=&
(\gamma^2_{mn}+(n\pi)^2)(\gamma^2_{mn'}+(n'\pi)^2)\nonumber\\
&&\{(1+\frac{\Gamma\beta_i}{2})\rm{fR3}+a^*_{mn}a_{mn'}\frac{{B_0^2}_e}{{B_0^2}_i}
(1+\frac{\Gamma \beta_e}{2})~\rm{fR4}\}\times\rm{fZ5} \nonumber\\
&&+((n\pi)^2(n'\pi)^2-(n\pi)^2(\gamma^2_{mn'}+
(n'\pi)^2)-(n'\pi)^2(\gamma^2_{mn}+(n\pi)^2))\nonumber\\
&&(\rm{fR3}+a^*_{mn}a_{mn'}\frac{{B_0^2}_e}{{B_0^2}_i}\rm{fR4})\times\rm{fZ5}\nonumber\\
&&+((n\pi)(n'\pi)(\rm{fR1}+a^*_{mn}a_{mn'}\frac{{B_0^2}_e}{{B_0^2}_i}\rm{fR2})\times\rm{fZ6},
\\
\frac{2}{B_0^2}_i W_{22}^{nn'}&=&
\gamma^2_{mn}\gamma^2_{mn'}(n\pi)(n'\pi) ( \rm{fR3} +a^*_{mn}a_{mn'}\frac{{B_0^2}_e}{{B_0^2}_i}\rm{fR4})
\times\rm{fZ5}\nonumber\\
&&+((n\pi)^2(n'\pi)^2(\rm{fR1}+a^*_{mn}a_{mn'}\frac{{B_0^2}_e}{{B_0^2}_i}\rm{fR2})\times\rm{fZ6},
\\
\frac{2}{B_0^2}_i W_{33}^{nn'}&=&
(n\pi)(n'\pi)(\gamma^2_{mn}+(n\pi)^2)(\gamma^2_{m n'}+(n'\pi)^2)\nonumber\\
&&(\rm{fR1}+a^*_{mn}a_{mn'}\frac{{B_0^2}_e}{{B_0^2}_i}\rm{fR2})
\times\rm{fZ5},
\\
\frac{2}{B_0^2}_i W_{12}^{nn'}&=&
(\gamma^2_{mn}+(n\pi)^2)\gamma^2_{mn'}(n'\pi)\nonumber\\
&& (\rm{fR3}+a^*_{mn}a_{mn'}\frac{{B_0^2}_e}{{B_0^2}_i}
\rm{fR4})\times\rm{fZ5},
\\
&&\frac{2}{B_0^2}_i W_{13}^{nn'}=0,
\\
&&\frac{2}{B_0^2}_i W_{23}^{nn'}=0.
\end{eqnarray}
In all formula, we scaled $z\to z/L$ and $r\to r/R$.
\end{document}